# Absence of dehydration due to superionic transition at Earth's core-mantle boundary

Yu He[1,2,#,*], Wei Zhang[3,#], Qingyang Hu[2], Shichuan Sun[4], Jiaqi Hu[1,5], Daohong Liu[1,5], Li Zhou[3], Lidong Dai[1], Duck Young Kim[2], Yun Liu[6,7], Heping Li[1], Ho-kwang Mao[2]

[1]Key Laboratory of High-Temperature and High-Pressure Study of the Earth's Interior, Institute of Geochemistry, Chinese Academy of Sciences, Guiyang 550081, Guizhou, China

[2]Center for High Pressure Science and Technology Advanced Research, Shanghai 201203, China

[3]School of Geography and Environmental Science (School of Karst Science), Guizhou Normal University, Guiyang 550025, China

[4]Asian School of the Environment, Nanyang Technological University, 50 Nanyang Avenue, Singapore 639798, Singapore

[5]University of Chinese Academy of Sciences, Beijing 100049, China

[6]State Key Laboratory of Ore Deposit Geochemistry, Institute of Geochemistry, Chinese Academy of Sciences, Guiyang 550081, China

[7]International Center for Planetary Science, College of Earth Sciences, Chengdu University of Technology, Chengdu 610059, China

*e-mail: heyu@mail.gyig.ac.cn

\# These authors contributed equally: Yu He, Wei Zhang

**The properties and stability of hydrous phases are key to unraveling the mysteries of the water cycle in Earth's interior. Under the deep lower mantle conditions, hydrous phases transition into a superionic state. However, the influence of the superionic effect on their stability and dehydration processes remains poorly understood. Using ab initio calculations and deep-learning potential molecular dynamics simulations, we discovered a doubly superionic transition in δ-AlOOH, characterized by the highly diffusive behavior of ionic hydrogen and aluminum within the oxygen sub-lattice. These highly diffusive elements contribute significant external entropy into the system, resulting in exceptional thermostability. Free energy calculations indicate that dehydration is energetically and kinetically unfavorable when water exists in a superionic state under core-mantle boundary (CMB) conditions. Consequently, water can accumulate in the deep lower mantle over Earth's history. This deep water reservoir plays a crucial role in the global deep water and hydrogen cycles.**

**Introduction**

Water, in the form of hydrous minerals, enters the mantle by subducting oceanic plates. During this process, most hydrous minerals dehydrate at high temperatures, releasing fluids that contribute to volcanism in island arcs[1-3]. The released water reacts with iron-rich olivine to produce natural hydrogen ($H_2$), which is increasingly recognized as an important renewable clean energy resource[4,5]. Some hydrous minerals, however, transform to high-density hydrous phases, such as phase H ($MgSiO_4H_2$)[6,7], δ-AlOOH[8-11], pyrite-type (py) $FeO_2H_x$ ($x≤1$)[12-15], and hydrous $SiO_2$[16-20]. In these hydrous phases, asymmetric O-H covalent bonds (hydroxyl) gradually transition to symmetric O-H-O ionic bonds as pressure increases[21-23]. This hydrogen-bond symmetrization enhances the stability of these hydrous phases, allowing them to coexist with bridgmanite and davemaoite[9,10,20] and enabling water to be transported into the lowermost mantle.

Intriguingly, water[24-28] and hydrous phases (py-$FeO_2H_x$, δ-AlOOH, and hydrous aluminous $SiO_2$)[29-32] undergo a transition to a superionic state under lower mantle conditions. This superionic transition results in liquid-like proton diffusion within the crystal lattice, leading to high electrical conductivities[30]. Furthermore, the presence of highly diffusive protons causes elastic softening and reduces shear wave velocities[33]. In superionic $H_2O$ ice, the additional entropy contributed by highly diffusive protons induces structural changes and increases the liquidus temperature[24-28]. However, the impact of the superionic effect on thermal stability and dehydration process of hydrous phase in Earth's interior is still unknown.

The stability and dehydration of hydrous phases are also critical for the generation and transportation of $H_2$ in deep lower mantle. Dehydrogenation has been observed during phase transition from ε-FeOOH ($CaCl_2$ structure) to py-$FeO_2H_x$ (x≤1) at ~70 GPa [14,15,34,35]. In addition, released water reacts with iron and iron oxides to produce py-$FeO_2H_x$[14,36,37], presenting significantly high density and low seismic velocities, comparable to the properties of ultra-low velocities zone (ULVZ) above CMB[15]. These reactions can also produce iron hydride, which may transport hydrogen into Earth's outer core[15,37,38]. It is interesting that hydrogen from deep mantle is found to be associated with high $^3$He content, suggesting primordial resource[4,39,40]. Alternatively, primordial hydrogen may be released form Earth's core through core-mantle reactions [4]. However, the relationship between subducted water and deep-released hydrogen is not established.

**Results and discussion**

**Double superionic transition in δ-AlOOH**

δ-AlOOH, with $CaCl_2$-type structure, is stable under deep lower mantle conditions[8-11]. As an end member, it forms $CaCl_2$-type AlOOH-$Mg_{0.5}Si_{0.5}$OOH-FeOOH and AlOOH-$SiO_2$ solid solution phases[6,9,18,20,41]. Additionally, the incorporation of Al dramatically increases the water solubility in stishovite and post-stishovite, making hydrous aluminous $SiO_2$ one of the dominate water hosts in the lower mantle[18,20]. Therefore, understanding the Al-O-H interaction is key to elucidating the properties of hydrous phases in the lower mantle.

Here, we conducted ab-initio molecular dynamics (AIMD) simulations on δ-AlOOH at 130 GPa and 2000-4000 K. Consistent with previous studies[32], we observed a superionic transition. Protons become highly diffusive at 3000 K (Fig. 1). Interestingly, we also detected $Al^{3+}$ migration along the b axis, which promotes proton diffusion in the same direction. This suggests that proton diffusion provides external migration site for $Al^{3+}$ (Supplementary Fig. S1). At 4000 K, the diffusion of $Al^{3+}$ is significant with mean squared displacement (MSD) increasing linearly with simulation time. Both $H^+$ and $Al^{3+}$ exhibit liquid-like diffusion within the oxygen sub-lattice, indicating a doubly superionic transition in δ-AlOOH (Fig. 1). The diffusion of $Al^{3+}$ reduces the anisotropy of $H^+$ diffusion, exhibited by the $H^+$ MSDs along different lattice directions (Supplementary Fig. S2).

Large-scaled molecular dynamic simulations based on deep-learning potentials (DP) were used to investigate diffusion of $H^+$ and $Al^{3+}$ in δ-AlOOH at ~70-130 GPa and ~1500-4000 K. The calculated densities are consistent with AIMD results (Supplementary Fig. S3). Diffusion coefficients were calculated using MSDs (Supplementary Fig. S4) and fitted with the Arrhenius equation (Fig. 2a). Based on typical ionic diffusion coefficients (~$10^{-11}$-$10^{-10}$ $m^2$ $s^{-1}$) in superionic materials[42], we estimated the superionic transition temperatures. The transition temperature for fast $H^+$ diffusion increases from ~2050 K to ~2270 K with rising pressure from 70 to 130 GPa, whereas $Al^{3+}$ diffusion becomes significant only at very high temperatures above ~3570 K. Using the diffusion coefficients, we calculated the ionic conductivities of δ-AlOOH (Fig. 2b). Although the contribution of $Al^{3+}$ to the total conductivity is marginal, the ionic conductivity of δ-AlOOH reaches ~600 S $m^{-1}$ at 4000 K. Under lower mantle conditions, the conductivities range from 1 to 10 S $m^{-1}$.

**The stability of δ-AlOOH**

In superionic δ-AlOOH, $H^+$ and $Al^{3+}$ are disordered and diffuse through interstitial sites within the lattice, leading to distinct thermodynamic properties. We calculated the

Gibbs free energies for solid δ-AlOOH, the superionic phase with diffusive $H^+$ (SI-I), the doubly superionic phase with diffusive $H^+$ and $Al^{3+}$ (SI-II), and the liquid phase using a two-step thermal integration (TI) method (see Methods). The resulting phase diagram is shown in Fig. 3. It reveals that superionic phases become energetically favorable at high temperatures, and ionic diffusion within the lattice stabilizes the solid structure, resulting in a high melting temperature of approximately 3800 K under core-mantle boundary (CMB) conditions (Supplementary Fig. S5). At pressures above ~140 GPa, the SI-II phase becomes unstable, and the SI-I phase transitions directly to the liquid phase at ~3900 K. This suggests that the predicted superionic transition based on diffusion coefficients (red stars) may be an artifact of not accounting for the superheating effect[33]. Here, we demonstrate that the two-step TI method provides accurate phase boundaries for solid-superionic-liquid transitions.

In addition, we investigated the dehydration of δ-AlOOH. We calculated the phase diagram of $H_2O$, and found superionic ice XVIII (face-centerd-cubic structure) is stable under CMB conditions. This result is consistent with previous experimental and computational studies (Supplementary Fig. S6)[24–28]. Consequently, rather than liquid $H_2O$, superionic $H_2O$ ice is more stable in the deep lower mantle. In this scenario, δ-AlOOH may dehydrate to produce superionic $H_2O$ ice XVIII and $Al_2O_3$ ($Ru_2O_3$-II structure). To evaluate this, we calculated the Gibbs free energy differences (ΔG) for this reaction at pressures of 80–140 GPa and temperatures of 2000–3600 K. The energy differences decrease with increasing temperature and pressure; however, they remain positive (Supplementary Fig. S7), indicating that the dehydration process is energetically unfavorable.

The dehydration mechanism of hydrous minerals is well understood at ambient pressure. At elevated temperature, $H^+$ becomes diffusive and reacts with hydroxyls to form water molecular at interstitial and/or dislocation sites within the lattice. Subsequently, these water molecules are released from the lattice, accompanied with a recrystallization process[43,44]. However, this mechanism is significantly different under deep lower mantle conditions. Hydroxyl groups transform into symmetric O-H-O configurations in hydrous phases (Supplementary Fig. S8) and $H_2O$ exists in a superionic state, which prevents its release from the lattice. Our simulations also indicate that oxygen is stable at lattice sites without significant diffusion, even at 4000 K (Fig. 1). As a result, dehydration under deep lower mantle conditions also becomes kinetically unfavorable as well.

High pressure experiments suggest the dehydration of δ-AlOOH occurs at temperatures approximately 2500 K under deep lower mantle conditions, based on the formation of $Al_2O_3$[10,11]. Conversely, recent studies have found that py-$FeO_2H_x$ (x < 1) and hydrous $SiO_2$ exhibit high stability. Py-$FeO_2H_x$ (x < 1) remains stable above 3000 K at 120-130 GPa[15,30], while aluminous $SiO_2$ containing 0.9-2.6 wt% $H_2O$ coexists with melts and remains stable even at temperature close to 4000 K under CMB conditions[20]. These results suggest that different hydrous phases process distinct stability field.

Our study indicates that superionic $H_2O$ ice XVIII is stable at 4000 K under CMB conditions. Hence, ice XVIII should be detectable in experiments if dehydration truly occurs. However, none of these experiments have observed an ice phase. This discrepancy suggests that the decomposition of δ-AlOOH may be due to reactions between δ-AlOOH and pressure mediums. It is quite likely that $H^+$ diffuses into pressure mediums, facilitating interdiffusion at high temperatures. We also note that in experiments where no pressure medium was used, highly stable hydrous phases were observed[15,19,20,30]. Additionally, δ-AlOOH was found to be more stable in experiments conducted without pressure mediums[11]. It indicates that partition of $H_2O$ may take place between different phases at high temperature due to fast proton diffusion. However, there is currently no evidence to suggest the complete decomposition of hydrous phases to generate a pure $H_2O$ phase under lower mantle conditions.

**Deep water cycle and dehydrogenation process**

Dehydration and dehydrogenation processes are crucial for understanding the water and hydrogen cycles in Earth's interior (Fig. 4). Our study demonstrates that these processes are strongly influenced by the state of water. Under upper mantle conditions, water exists in liquid form, and hydrous phases dehydrate during subduction. The released water promotes serpentinization and can oxidize ferrous ions in minerals, generating natural $H_2$. In contrast, under lower mantle conditions, the symmetrization (ionization) of O-H bonds[21–23] and the superionic transition[30–32] significantly enhance the stability of hydrous phases. Notably, $H_2O$ exists in a superionic state even under core-mantle boundary (CMB) conditions (Fig. 3), making dehydration both energetically and kinetically unfavorable. Due to the high stability of hydrous phases, some ancient water may have been locked in the deep lower mantle throughout Earth's history.

On the other hand, experimental and computational studies suggest partial dehydrogenation from ε-FeOOH during the formation of py-$FeO_2H_x$ ($x < 1$) at pressures exceeding ~70 GPa[14,15,34,35]. Under lowermost mantle, $Al^{3+}$ ions become highly diffusive in δ-AlOOH, promoting interdiffusion between $Al^{3+}$ and $Fe^{3+}$. The presence of $Fe^{3+}$ in hydrous phases leads to the formation of py-$FeO_2H_x$ and gradual release of $H_2$[41]. Alternatively, water-bearing phases are transported to the lowermost mantle, where they react with Fe from the outer core to produce $H_2$, iron hydride, and iron oxides[364,37,46]. The released $H_2$ can be transported back to Earth's surface by mantle plumes connecting large low shear velocity provinces (LLSVPs) to locations of hotspot volcanism (Fig. 4)[47]. Interestingly, hydrogen isotopic anomalies have been observed in natural $H_2$[4] and ocean-island basalts (OIB)[39,40], accompanied by the enrichment of $^3He$, a primordial isotope incorporated into Earth during its accretion. The superionic effect facilitates the mixing of subducted hydrogen with primordial hydrogen, which is characterized by a low deuterium-to-hydrogen (D/H) ratio. This primordial hydrogen may have been introduced into the deep Earth through an ingassing process from the atmosphere during the early stages of Earth's formation[48]. Coincidentally, helium (He) can also be encapsulated within the same structures ($CaCl_2$ and pyrite structures) as hydrous phases under lower mantle conditions[49,50]. The dehydrogenation process may also promote the leakage of He from the lattice, resulting in the enriched $^3He$ signatures observed in volcanic/magmatic natural $H_2$ and OIB.

LLSVPs above CMB, presenting low velocity[51] and sharp boundary[52] being distinct from surrounding mantle, are considered to be caused by thermochemical heterogeneity constructing by primordial material[53] and/or the accumulation of subducted oceanic curst[54]. The seismic velocity features of LLSVPs can also be explained by the presence of hydrous phases[55] or hydrous bridgmanite[56]. Therefore, both geophysical and geochemical observations suggest that LLSVPs may act as sinks for ancient water in the lower mantle, with the dehydrogenation process leading to the release of $H_2$ and He with primordial signature. The ancient water reservoir located deep within Earth could be crucial to reveal the mysteries of water and hydrogen deep cycling from the early stage of Earth's history to the present. Moreover, the reservoir may also function as a large hydrogen generator, supplying unlimited clean and carbon-free fuel from the deepest mantle to the surface.

## Methods

### *Ab-initio* molecular dynamics (AIMD) simulations on superionic transition in δ-AlOOH

AIMD simulations are widely utilized to investigate superionic transitions in Earth and planetary interiors. This study employs AIMD simulations based on Density Functional Theory (DFT)[57] using the Vienna Ab Initio Simulation Package (VASP)[58]. We used atomic potentials generated using the projector augmented-wave method (PAW)[59] within the generalised gradient approximations (GGA)[60]. The energy cut-off was set to 600 eV and Brillouin zone sampling was performed at the Γ point. A 2×2×4 supercell (128 atoms) of δ-AlOOH with $CaCl_2$ structure was used for AIMD simulations. The hydrostatic structures at 130 GPa and 2000-4000 K were obtained by conducting simulations under *NPT* (*N*, number of particles; *P*, pressure; *T*, temperature) ensemble with a time step of 1 fs for total 20 ps. Then the self-diffusion behavior was calculated using canonical ensemble (*NVT*) with a time step of 1 fs for total 20 ps. Langevin and

Nosé thermostats were used for *NPT* and *NVT* simulations, respectively. The time averaged mean square displacements (MSDs) of $Al^{3+}$, $H^+$, and $O^{2-}$ were calculated using the atomic configurations from each simulation time step using equation (1):

$$\langle[\vec{r}(t)]^2\rangle = \frac{1}{N}\sum_{i=1}^{N}\langle[\vec{r_i}(t+t_0) - \vec{r_i}(t_0)]^2\rangle \quad (1),$$

where $\vec{r_i}(t)$ is the displacement of the ith ion at time *t*, and *N* is the total number of protons in the system. In practice, *D* is obtained by a linear fit to the time dependence of the average MSDs.

$$D = \lim_{0\to\infty}\left[\frac{1}{2dt}\langle[\vec{r}(t)]^2\rangle\right] \quad (2).$$

In superionic state, part of ions ($H^+$ and $Al^{3+}$ in this study) in the lattice presents significant self-diffusion with MSDs increasing with simulation time, and other part of ions ($O^{2-}$) vibrates at their lattice sites with insignificant MSDs increasement during the simulation.

**Deep-potential training method and DFT calculation**

The pretraining of the attention-based model (DPA-1)[61], implemented in the DeePMD-kit package (v2.2.7)[61-64], was employed for deep learning potential model training. An active learning procedure using the deep potential generator (DP-GEN) package[65] efficiently generated training datasets through an iterative exploration process. Six different structures, including δ-AlOOH, AlOOH (liquid), $Al_2O_3$ ($Ru_2O_3$-II structure), Ice X (body-centered cubic structure, bcc), Ice XVIII (face-centered cubic structure, fcc), and liquid water, were adopted in the DP-GEN iterations to calculate their energies, forces, and stresses for the training database. These iterations were conducted at temperatures ranging from 1000 to 5000 K and pressures from 60 to 160 GPa. In total, 10,006 configurations were selected for model training, and 388 configurations were selected for model testing. During DP-GEN iterations, the cut-off radius was set to 6 Å, and the inverse distance 1/*r* decayed smoothly from 0.5 Å to 6 Å to eliminate discontinuity introduced by the cut-off. The sizes of the embedding network and fitting network were {25, 50, 100} and {240, 240, 240}, respectively. The productive model used the same training parameters as the DP-GEN iterations, except that the smooth cut-off was adjusted from 0.5 Å to 10 Å to capture long-range interactions.

All DFT labeling for deep potential training was performed using ABACUS (version 3.6.0) with a Linear Combination of Atomic Orbitals (LCAO) basis set[66,67]. The calculations utilized double-zeta plus polarization (DZP) numerical atomic orbitals and norm-conserving pseudopotentials (SG15 ONCV)[68]. The PBE functional[58] was applied, along with Grimme's D3 dispersion correction[69]. Atom counts for the DFT labeling were 64 for AlOOH, 80 for $Al_2O_3$, 162 for Ice X, 96 for Ice XVIII, and 192 for liquid water. The kinetic energy cutoff was set at 100 Ry, with an SCF convergence threshold for the density error of $1\times10^{-6}$ Ry, and a K-spacing of 0.2.

Additionally, four extra AIMD simulations were conducted on the AlOOH system at 1500 K, 2500 K, 3500 K, and 4000 K using the *NVT* ensemble. The K-spacing was set to 0.2, with a time step of 0.3 fs for thermalization. Each system was composed of 192 atoms. The initial configurations were obtained at 100 GPa at the respective temperatures of interest.

The accuracy of the DP model was initially evaluated by comparing its predictions to DFT calculations. Supplementary Fig. S9 and Fig. S10 show the comparison of force and energy for the test data sets. The overall prediction error in force values is 333 meV Å$^{-1}$. The energy prediction errors are 8.08, 2.59, and 7.17 meV atom$^{-1}$ for the AlOOH, $Al_2O_3$, and ice (ice X, ice XVIII, and liquid) systems, respectively. The comparison of radial distribution functions (RDFs) between deep potential molecular dynamic (DPMD) and AIMD, illustrated in Supplementary Fig. S11, shows a close agreement. This demonstrates the convergence of the DP model, indicating that the trained DP model achieves an accuracy comparable to that of DFT.

**DPMD simulation and Gibbs free energy calculation**

Using the DP model, we conducted DPMD simulations using the Large-scale

Atomic/Molecular Massively Parallel Simulator (LAMMPS) package[70]. Temperature and pressure were controlled using a Langevin thermostat[71] and a Nosé–Hoover barostat[72], respectively. A time step of 0.3 fs was employed. The volume of the simulation cells was determined from 12 ps of DPMD simulations within *NPT* ensemble at the desired temperature and pressure, using the average box size from the latter half of the simulation. In δ-AlOOH, the calculated *P-V-T* relations present good consistence with the values calculated using AIMD simulation (Supplementary Fig. 3S). We calculated Gibbs free energies of ice and AlOOH (solid, superionic, and liquid phases) at pressures ranging from 60 to 160 GPa and temperatures between 1800 and 4600 K using the nonequilibrium thermodynamic integration (NeTI) method[73-76].

**Acknowledgements**

We acknowledge Dr. Maurice de Koning for discussion on NeTI method. The project is supported by National Key Research and Development Program of China, (Grant No. 2024YFF0807500). We acknowledge the support of the National Natural Science Foundation of China (42350002, 42303071), the CAS Youth Interdisciplinary Team (JCTD-2022-16), and the Youth Innovation Promotion Association of CAS (2020394). This study was also supported by the Guizhou Provincial 2020 Science and Technology Subsidies (No. GZ2020SIG). Numerical computations were performed at the Shanghai Supercomputer Center and National Supercomputer Center in Guangzhou. The computing resources used in this study were also provided by the Bohrium Cloud Platform (https://bohrium.dp.tech), which is supported by DP Technology.

**Author contributions**

Y.H. and W.Z. contributed equally to this work, conducted the calculations, analyzed the data and wrote the original manuscript. Y.H., Q.H., D.Y.K. and H.-k.M. initiated and designed the project. Y.H. and J.H. conducted DFT calculations. W.Z., S.S. and D.L. constructed deep potentials and performed DPMD simulations. Y.H., L.D., L.Z. and H.L. discussed the geophysical implications. Y.H., Y.L. and Q.H. discussed the geochemical implications. All authors discussed the data interpretation and commented on the manuscript.

## Declaration of Competing Interest

The authors declare that they have no known competing financial interests.

## Code availability

The Vienna Ab initio Simulation Package is proprietary software available for purchase at https://www.vasp.at/. LAMMPS software is available at https://www.lammps.org/. DeePMD-kit code is available at https://github.com/deepmodeling/deepmd-kit. ABACUS software is available at https://abacus.ustc.edu.cn/main.htm.

# Reference:

1. Williams, Q. & Hemley, R. Hydrogen in the deep Earth. *Annu. Rev. Earth Planet. Sci.* **29**, 365–418 (2001).
2. Ohtani, E. The role of water in Earth's mantle. *Natl Sci. Rev.* **7**, 224–232 (2020).
3. Ohtani, E. Hydration and dehydration in Earth's interior. *Annu. Rev. Earth Planet. Sci.* **49**, 253–278 (2021).
4. Milkov, A. V. Molecular hydrogen in surface and subsurface natural gases: Abundance, origins and ideas for deliberate exploration. *Earth-Sci. Rev.* **230**, 104063 (2022).
5. Hand, E. Hidden hydrogen, does Earth hold vast stores of a renewable, carbon-free fuel? *Science* **379**, 630-636 (2023).
6. Nishi, M. et al. Stability of hydrous silicate at high pressures and water transport to the deep lower mantle. *Nat. Geosci.* **7**, 224–227 (2014).
7. Ohtani, E., Amaike, Y., Kamada, S., Sakamaki, T. & Hirao, N. Stability of hydrous phase H $MgSiO_4H_2$ under lower mantle conditions. *Geophys. Res. Lett.* **41**, 8283–8287 (2014).
8. Sano, A. Ohtani, E. & Kondo, T. et al. Aluminous hydrous mineral δ-AlOOH as a carrier of hydrogen into the core–mantle boundary. *Geophys. Res. Lett.* **35**, 303 (2008).
9. Ohira, I. et al. Stability of a hydrous δ-phase, $AlOOH–MgSiO_2(OH)_2$, and a mechanism for water transport into the base of lower mantle. *Earth Planet. Sci. Lett.* **401**, 12–17 (2014).
10. Duan, Y. et al. Phase stability and thermal equation of state of δ-AlOOH: implication for water transportation to the deep lower mantle. *Earth Planet. Sci. Lett.* **494**, 92–98 (2018).
11. Piet, H. et al. Dehydration of δ-AlOOH in Earth's Deep Lower Mantle. *Minerals* **10**, 384 (2020).
12. Hu, Q. et al. $FeO_2$ and FeOOH under deep lower-mantle conditions and Earth's oxygen–hydrogen cycles. *Nature* **534**, 241–244 (2016).
13. Nishi, M. et al. The pyrite-type high-pressure form of FeOOH. *Nature* **547**, 205–208 (2017).
14. Hu, Q. et al. Dehydrogenation of goethite in Earth's deep lower mantle. *Proc. Natl. Acad. Sci. USA* **114**, 1498-501 (2017).
15. Liu, J. et al. Hydrogen-bearing iron peroxide and the origin of ultralow-velocity zones. *Nature* **551**, 494–497 (2017).
16. Lin, Y., Hu, Q., Meng, Y., Walter, M. & Mao, H. K. Evidence for the stability of ultrahydrous stishovite in Earth's lower mantle. *Proc. Natl Acad. Sci. USA* **117**, 184–189 (2020).
17. Nisr, C. et al. Large $H_2O$ solubility in dense silica and its implications for the interiors of water-rich planets. *Proc. Natl Acad. Sci. USA* **117**, 9747–9754 (2020).
18. Ishii, T. et al. Superhydrous aluminous silica phases as major water hosts in high-temperature lower mantle. *Proc. Natl Acad. Sci. USA* **119**, e2211243119 (2022).
19. Lin, Y. et al. Hydrous $SiO_2$ in subducted oceanic crust and $H_2O$ transport to the core–mantle boundary. *Earth Planet. Sci. Lett.* **594**, 117708 (2022).
20. Tautsumi, Y. et al. Retention of water in subducted slabs under core–mantle boundary conditions. *Nat. Geosci.* **17**, 697-704 (2024).
21. Tsuchiya, J. Tsuchiya, T. & Tsuneyuki, S. et al. First principles calculation of a high-pressure hydrous phase, δ-AlOOH. *Geophys. Res. Lett.* **29**, 1909 (2002).
22. Panero, W. R. & Stixrude, L. P. Hydrogen incorporation in stishovite at high pressure and symmetric hydrogen bonding in δ-AlOOH. *Earth Planet. Sci. Lett.* **221**, 421-431 (2004).

23. Meier, T. et al. Structural independence of hydrogen-bond symmetrisation dynamics at extreme pressure conditions. *Nat. Commun.* **13**, 3042 (2022).
24. Millot M. et al. Nanosecond X-ray diffraction of shock-compressed superionic water ice. *Nature* **569**, 251-255 (2019).
25. Prakapenka V. B., Holtgrewe, N. Lobanov, S. S. & Goncharov A. F. Structure and properties of two superionic ice phases. *Nat. Phys.* **17**, 1233-1238 (2021).
26. Weck, G. et al. Evidence and Stability Field of fcc Superionic Water Ice Using Static Compression. *Phys. Rev. Lett.* **128**, 165701 (2022).
27. Husband, R. J. et al. Phase transition kinetics of superionic $H_2O$ ice phases revealed by Megahertz X-ray freeelectron laser-heating experiments. *Nat. Commun.* **15**, 8256 (2024).
28. Cheng, B., Bethkenhagen, M., Pickard, C. J. & Hamel, S. Phase behaviours of superionic water at planetary conditions. *Nat. Phys.* **17**, 1228-1232 (2021).
29. Umemoto, K., Kawamura, K., Hirose, K. & Wentzcovitch, R. M. Post-stishovite transition in hydrous aluminous $SiO_2$. *Phys. Earth Planet. Inter.* **255**, 18–26 (2016).
30. Hou, M. et al. Superionic iron oxide–hydroxide in Earth's deep mantle. *Nat. Geosci.* **14**, 174–178 (2021).
31. Li, J. et al. Silica-water superstructure and one-dimensional superionic conduit in Earth's mantle. *Sci. Adv.* **9**, eadh3784 (2023).
32. Luo, C., Sun, Y. & Wentzcovitch, R. M. Probing the state of hydrogen in δ-AlOOH at mantle conditions with machine learning potential. *Phys. Rev. Res.* **6**, 013292 (2024).
33. He, Y. et al. Superionic iron alloys and their seismic velocities in Earth's inner core. *Nature* **602**, 258-262 (2022).
34. Zhu, S. et al., Hydrogen-Bond Symmetrization Breakdown and Dehydrogenation Mechanism of $FeO_2H$ at High Pressure. *J. Am. Chem. Soc.* **139**, 12139-12132 (2017).
35. Tang, R. et al., Chemistry and P-V-T equation of state of $FeO_2H_x$ at the base of Earth's lower mantle and their geophysical implications. *Sci. Bullet.* **66**, 1954-1958 (2021).
36. Yuan L. et al., Chemical Reactions Between Fe and $H_2O$ up to Megabar Pressures and Implications for Water Storage in the Earth's Mantle and Core. *Geophys. Res. Lett.* **45**, 1330–1338 (2018).
37. Mao, H.-K. et al. When water meets iron at Earth's core–mantle boundary. *Natl Sci. Rev.* **4**, 870–878 (2017).
38. He, Y. et al. The stability of $FeH_x$ and hydrogen transport at Earth's core mantle boundary. *Sci. Bullet.* **68**, 1567–1573 (2023).
39. Hallis, L. J. et al. Evidence for primordial water in Earth's deep mantle. *Science* **350**, 795-797 (2015).
40. Loewen, M. W. et al. Hydrogen isotopes in high 3He/4He submarine basalts: Primordial vs. recycled water and the veil of mantle enrichment. *Earth Planet. Sci. Lett.* **508**, 62-73 (2019).
41. Yuan, H. et al. Stability of Fe-bearing hydrous phases and element partitioning in the system $MgO–Al_2O_3–Fe_2O_3–SiO_2–H_2O$ in Earth's lowermost mantle. *Earth Planet. Sci. Lett.* **524**, 115714 (2019).
42. Hull, S. Superionics: crystal structures and conduction processes. *Rep. Prog. Phys.* **67**, 1233-1314 (2004).
43. Brindley, G. W. & Hayami, R. Kinetics and mechanism of dehydration and recrystallization of serpentine. *Clays and Clay Minerals*, **12**, 35–54 (1964).

44. Maiti, G. C. & Freund F. Dehydration-related proton conductivity in kaolinite. *Clay Minerals.* **16**, 395-413 (1981).
45. Brown, J. M. & Shankland, T. J. Thermodynamic parameters in the Earth as determined from seismic profiles. *Geophys. J Roy. Astron. Soc.* **66**, 579-96 (1981).
46. Terasaki, H. et al., Stability of Fe-Ni hydride after the reaction between Fe-Ni alloy and hydrous phase (d-AlOOH) up to 1.2 Mbar: possibility of H contribution to the core density deficit. *Phys. Earth Planet Inter.* **194–195**, 18–24 (2012).
47. Koppers, A. A. Mantle plumes and their role in Earth processes. *Nat. Rev. Earth Environ.* **2**, 382–401 (2021).
48. Olson, P. & Sharp, Z. D. Hydrogen and helium ingassing during terrestrial planet accretion. *Earth Planet. Sci. Lett.* **498**, 418-426 (2018).
49. Liu, H. et al. A first-principles study of the structural, electronic and elastic properties of the $FeO_2$–$FeO_2He$ system under high pressure. *Phys. Chem. Chem. Phys.* **25**, 20225 (2023).
50. Ding, S. et al. Formation of solid $SiO_2He$ compound at high pressure and high temperature. *Phys. Rev. B* **106**, 024102 (2022).
51. Garnero, E. J., McNamara A. K. & Shim S.-H. Continent-sized anomalous zones with low seismic velocity at the base of Earth's mantle. *Nat. Geosci.* **9**, 481-489 (2016).
52. Ni, S. D., et al., Sharp sides to the African superplume. *Science* **296**, 1850-1852 (2002).
53. Labrosse, S., Hernlund, J.W. & Coltice, N. A crystallizing dense magma ocean at the base of the Earth's mantle. *Nature* **450**, 866–869 (2007).
54. Tan, E. & Gurnis, M. Metastable superplumes and mantle compressibility. *Geophy. Res. Lett.* **32**, L20307 (2005).
55. Mashino, I., Murakimi, M. & Ohtani E. Sound velocities of δ-AlOOH up to core-mantle boundary pressures with implications for the seismic anomalies in the deep mantle. *J. Geophys. Res. Solid Earth* **121**, 595–609 (2016).
56. Jiang, J. & Zhang, F. Theoretical studies on the hydrous lower mantle and D" layer minerals. *Earth Planet. Sci. Lett.* **525**, 115753 (2019).
57. Kohn, W. & Sham, Self-consistent equations including exchange and correlation effects. L. J. *Phys. Rev.* **140**, A1133 (1965).
58. Kresse, G. Efficient iterative schemes for ab initio total-energy calculations using a plane-wave basis set. *Phys. Rev. B* **54**, 11169–11186 (1996).
59. Blöchl, P. E. Projector augmented-wave method. *Phys. Rev. B* **50**, 17953–17979 (1994).
60. Perdew, J. P., Burke, K. & Ernzerhof, M. Generalized gradient approximation made simple. *Phys. Rev. Lett.* **77**, 3865–3868 (1996).
61. Zhang, D. et al. Pretraining of attention-based deep learning potential model for molecular simulation. *npj Comput. Mater.* **10**, 94 (2024).
62. Wang, H., Zhang, L., Han, J. & Weinan, E. DeePMD-kit: A deep learning package for many-body potential energy representation and molecular dynamics. *Comput. Phys. Commun.* **228**, 178-184 (2018).
63. Zhang, L., Han, J., Wang, H., Car, R. & Weinan, E. Deep potential molecular dynamics: a scalable model with the accuracy of quantum mechanics. *Phys. Rev. Lett.* **120**, 143001 (2018).
64. Zeng, J. et al. DeePMD-kit v2: A software package for deep potential models. *J. Chem. Phys.* **159** (2023).

65. Zhang, Y. et al. DP-GEN: A concurrent learning platform for the generation of reliable deep learning based potential energy models. *Comput. Phys. Commun.* **253**, 107206 (2020).
66. Chen, M., Guo, G. C. & He, L. Systematically improvable optimized atomic basis sets for ab initio calculations. *J. Phys.: Condens. Matter* **22**, 445501 (2010).
67. Li, P. et al. Large-scale ab initio simulations based on systematically improvable atomic basis. *Comput. Mater. Sci.* **112**, 503-517 (2016).
68. Hamann, D. R. Optimized norm-conserving Vanderbilt pseudopotentials. *Phys. Rev. B* **88**, 085117 (2013).
69. Grimme, S., Antony, J., Ehrlich, S. & Krieg, H. A consistent and accurate ab initio parametrization of density functional dispersion correction (DFT-D) for the 94 elements H-Pu. *J. Chem. Phys.* **132**, 154104 (2010).
70. Plimpton, S. Fast Parallel Algorithms for Short-Range Molecular Dynamics. *J. Comput. Phys.* **117**, 1-19 (1995).
71. Schneider, T. & Stoll, E. Molecular-dynamics study of a three-dimensional one-component model for distortive phase transitions. *Phys. Rev. B* **17**, 1302-1322 (1978).
72. Hoover, W. G. Canonical dynamics: Equilibrium phase-space distributions. *Phys. Rev. A* **31**, 1695-1697 (1985).
73. Freitas, R., Asta, M. & de Koning, M. Nonequilibrium free-energy calculation of solids using LAMMPS. *Comput. Mater. Sci.* **112**, 333-341 (2016).
74. Paula Leite, R. & de Koning, M. Nonequilibrium free-energy calculations of fluids using LAMMPS. *Computational Materials Science* **159**, 316-326 (2019).
75. Fidalgo Cândido, V., Matusalem, F. & de Koning, M. Melting conditions and entropies of superionic water ice: Free-energy calculations based on hybrid solid/liquid reference systems. *J. Chem. Phys.* **158** (2023).
76. Cajahuaringa, S. & Antonelli, A. Non-equilibrium free-energy calculation of phase-boundaries using LAMMPS. *Comput. Mater. Sci.* **207**, 111275 (2022).

# Figure captions:

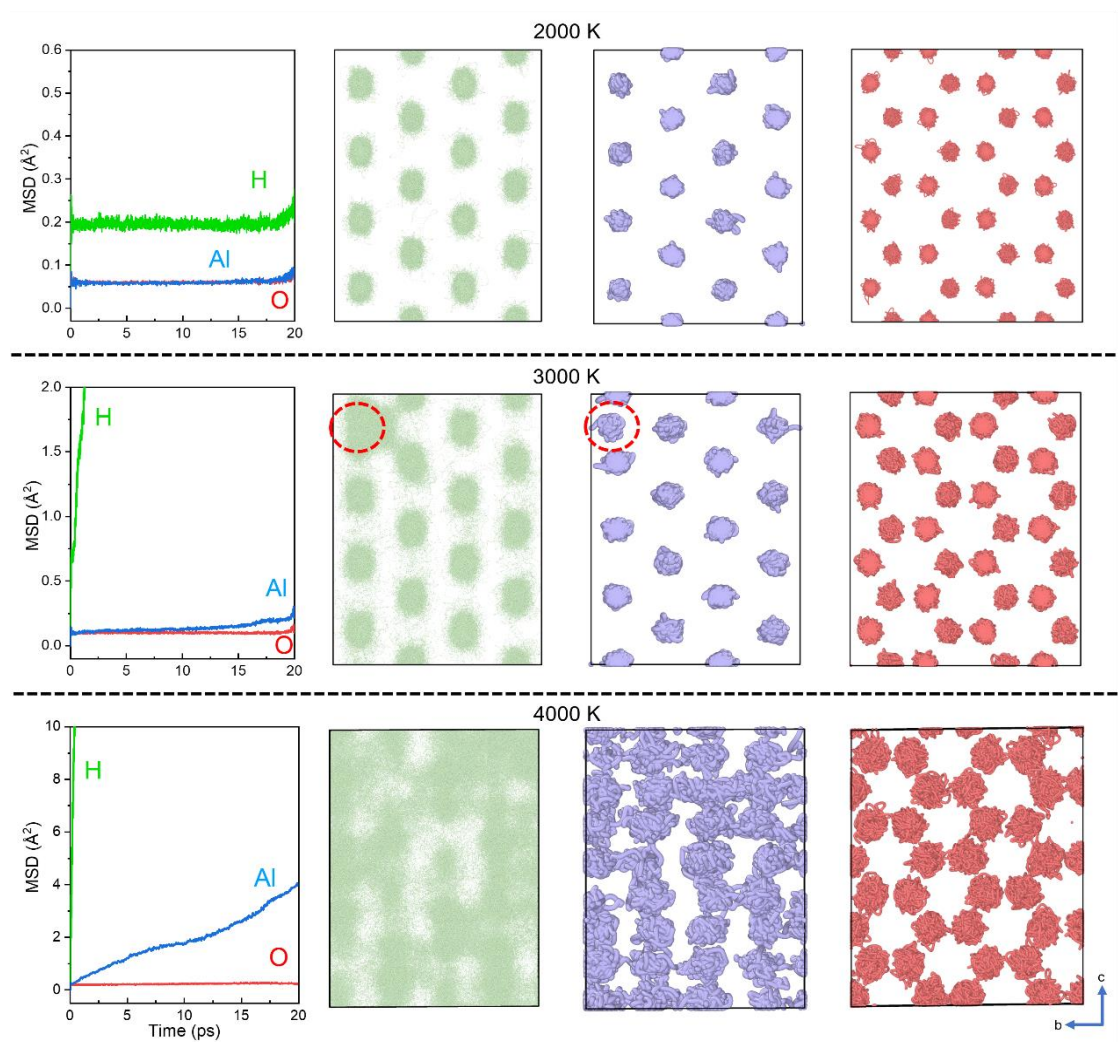

**Fig. 1 | The mean square displacements (MSDs) and trajectories of H, Al, and O ions in δ-AlOOH upon double superionic transition with increasing temperature from 2000 to 4000 K at 130 GPa.** The MSDs of H, Al, and O ions are exhibited in the left side with green, blue, and red curves. The linearly increase of MSDs with simulation time at 3000 K for $H^+$ and 4000 K for $Al^{3+}$ indicating fast ionic diffusion. The trajectories of $H^+$, $Al^{3+}$, and $O^{2-}$ are shown with green, blue, and red spheres on the right side. The disordered distribution of $H^+$ and $Al^{3+}$ are observed above 3000 K indicating a superionic transition. The migration of $Al^{3+}$ are also observed at 3000 K noted with red dashed circles resulting in enhanced proton diffusion along b axis.

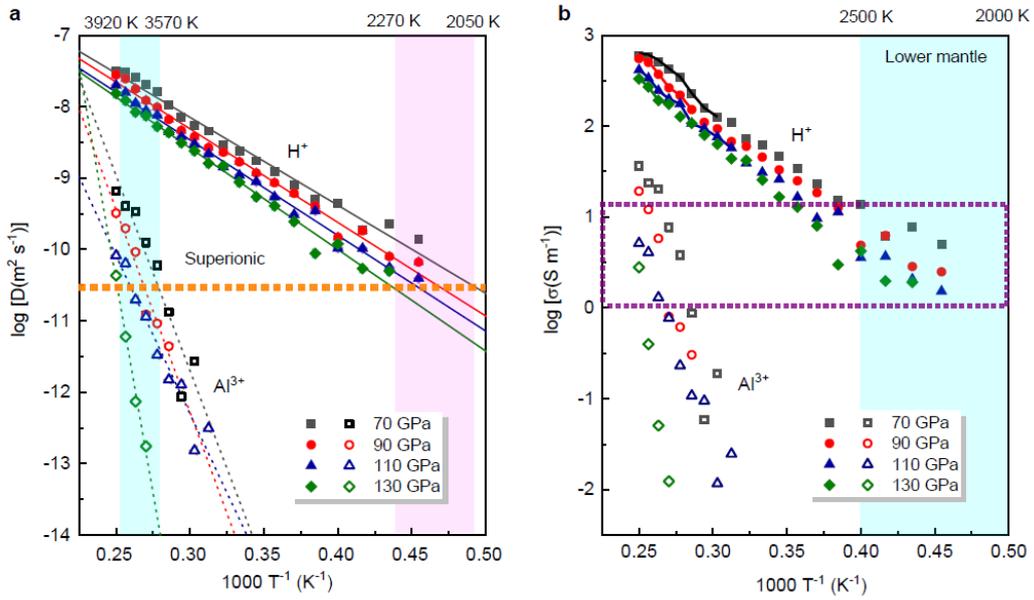

**Fig. 2 | Diffusion coefficients and ionic conductivities of δ-AlOOH at 70-130 GPa and 2000-4000 K.** The calculated (DPMD) diffusion coefficients (a) and ionic conductivities (b) of $H^+$ and $Al^{3+}$ are shown with filled and empty symbols respectively. The diffusion coefficients are fitted with an Arrhenius equation (solid and dashed lines). The dashed orange line represents the typical diffusion coefficient for superionic matters. Based on this line, the predicted superionic transition temperatures are shown with pink and clay region. The total ionic conductivities of δ-AlOOH are shown with solid curves suggesting the contribution of $Al^{3+}$ is insignificant. The conductivities of δ-AlOOH under lower mantle conditions (clay region) are included by dashed purple rectangle.

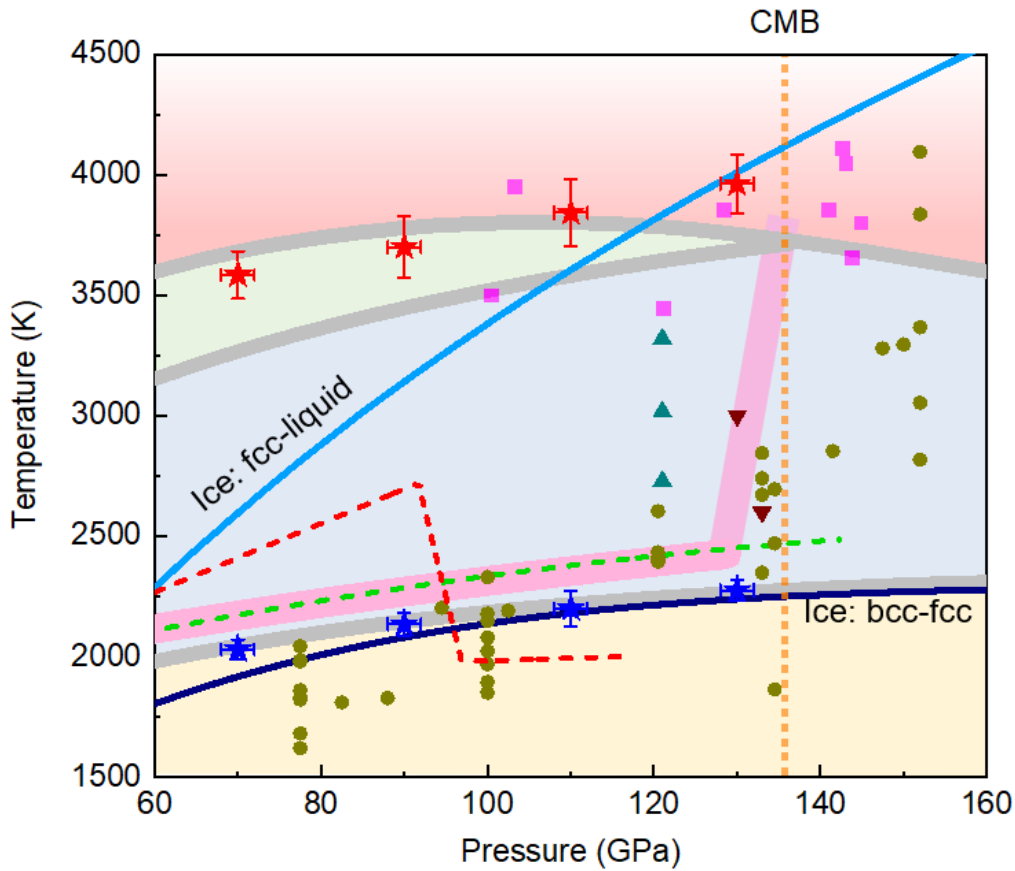

**Fig. 3 | Phase diagram of δ-AlOOH in comparison with H₂O at 60-160 GPa and 1500-4500 K.** Phase transition from solid (yellow) to SI-I (blue) to SI-II (green) and to liquid (red). Green[10] and red[11] dashed curves present the instability of δ-AlOOH suggested by high pressure experiments. The stability field of hydrous $SiO_2$, pyrite $FeO_2H$, and hydrous aluminous $SiO_2$ are shown with solid brown circles[19], cyan triangles[30], dark red inverted-triangles[15], and magenta squares[20], respectively. The P-V of highly $H^+$ and $Al^{3+}$ diffusion (superionic transition) is shown with open blue and red starts (with errors from the linearly fitted diffusion coefficients). Phase boundaries of ice X (*bcc*) to ice XVIII (*fcc*), and ice XVIII (*fcc*) to liquid are exhibited with dark blue and light blue curves. Lower mantle normal geotherm is shown with a pink belt[45]. The geotherm increases dramatically to over 3500-3800 K at CMB regions (orange dashed line).

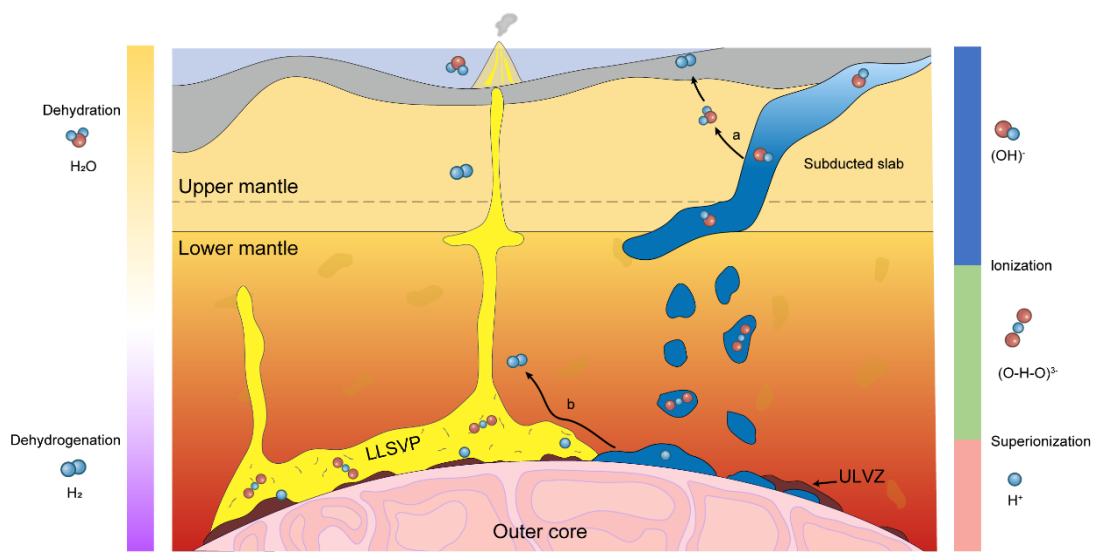

**Fig. 4 | Deep water reservoir and circulation.** Water can be transported into Earth's interior by subduction slabs in the form of hydroxyl. In upper mantle, the dehydration of hydrous minerals leading to serpentinization and release $H_2$. Covalent O-H bonding present symmetric ionization under lower mantle conditions. At CMB region, superionic transition results in high stability of hydrous phases and make lower mantle a potential water reservoir. Reactions between hydrous phases and iron oxides produce deep-seated $H_2$, which can be transport back to Earth's surface by mantle plume.

**Supplementary Information Figures and Tables**

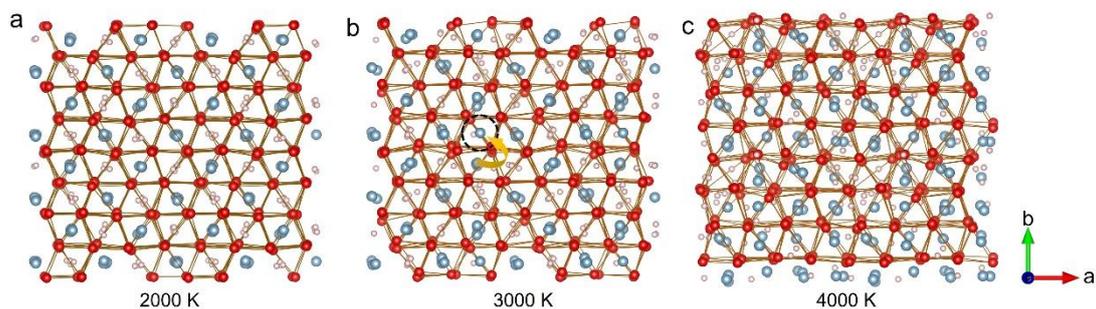

**Fig. S1.** The lattice structure of δ-AlOOH after 20 ps AIMD simulations at 130 GPa and temperatures of (a) 2000, (b) 3000, and (c) 4000 K. H, Al, and O are shown with light pink, light blue, and red spheres. The O sub-lattices (dark red lines) are stable after simulations. The yellow arrow (b) notes a $Al^{3+}$ migration along b axis, observed in the simulation at 3000 K from the Al-site to H-site (dashed black circle) in δ-AlOOH.

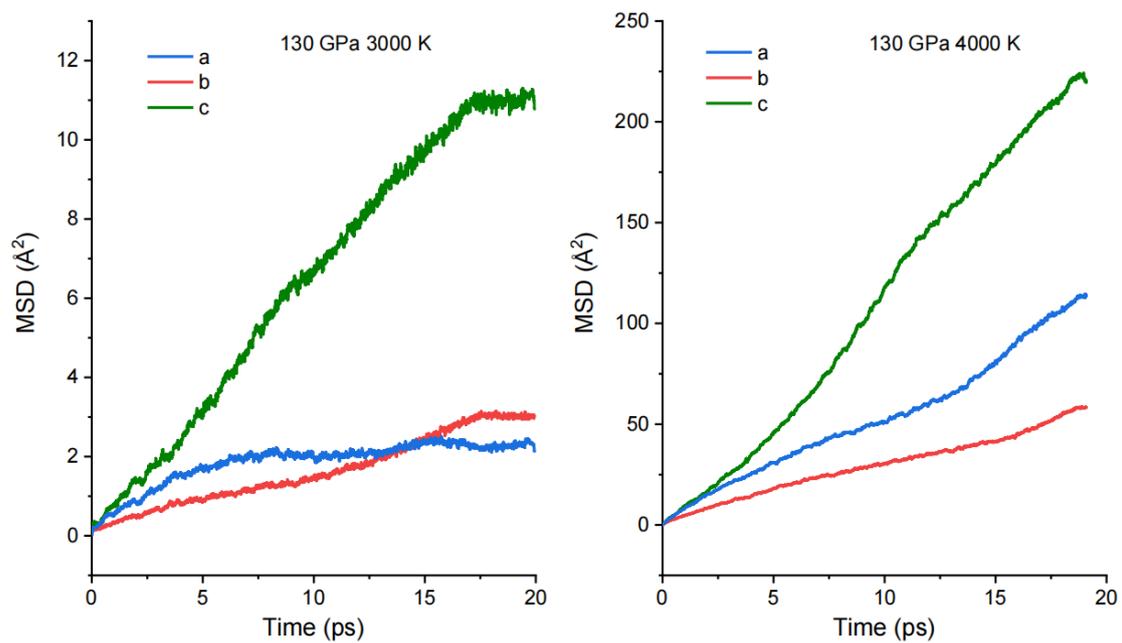

**Fig. S2.** MSDs of H$^+$ along a, b, and c lattice directions in δ-AlOOH at 130 GPa, and 3000 and 4000 K.

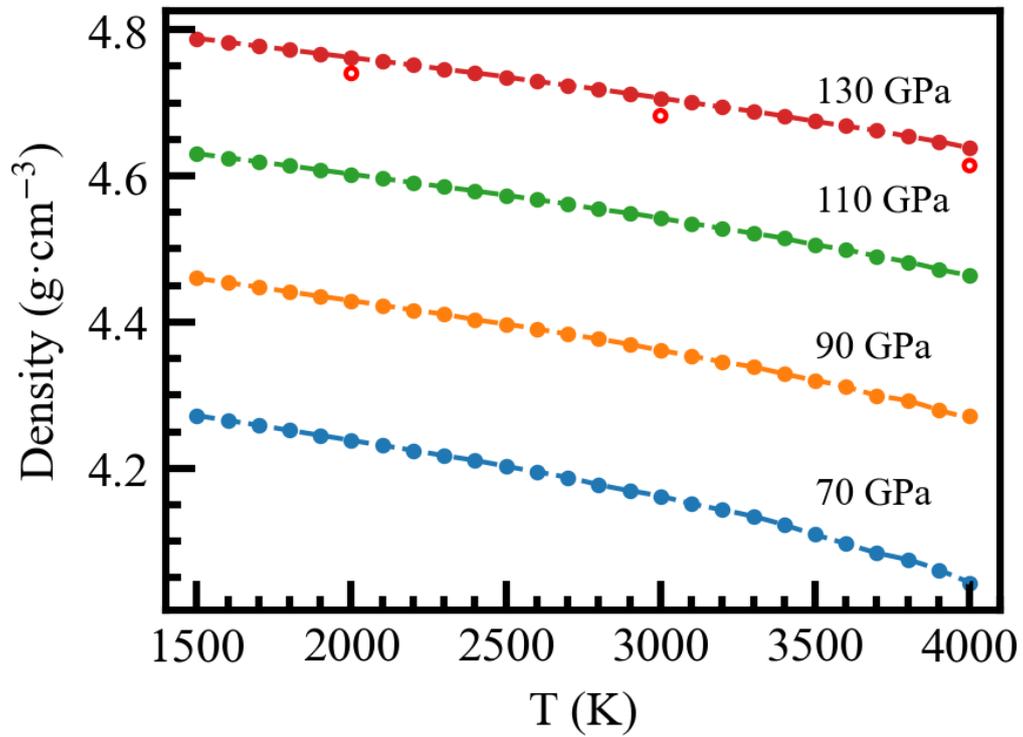

**Fig. S3.** Calculated density ($\rho$)-pressure ($P$)-temperature ($T$) rations of δ-AlOOH using DPMD method at 70-130 GPa and 1500-4000 K. The data at 130 GPa is compared with the results computed using AIMD simulations (empty red cycles).

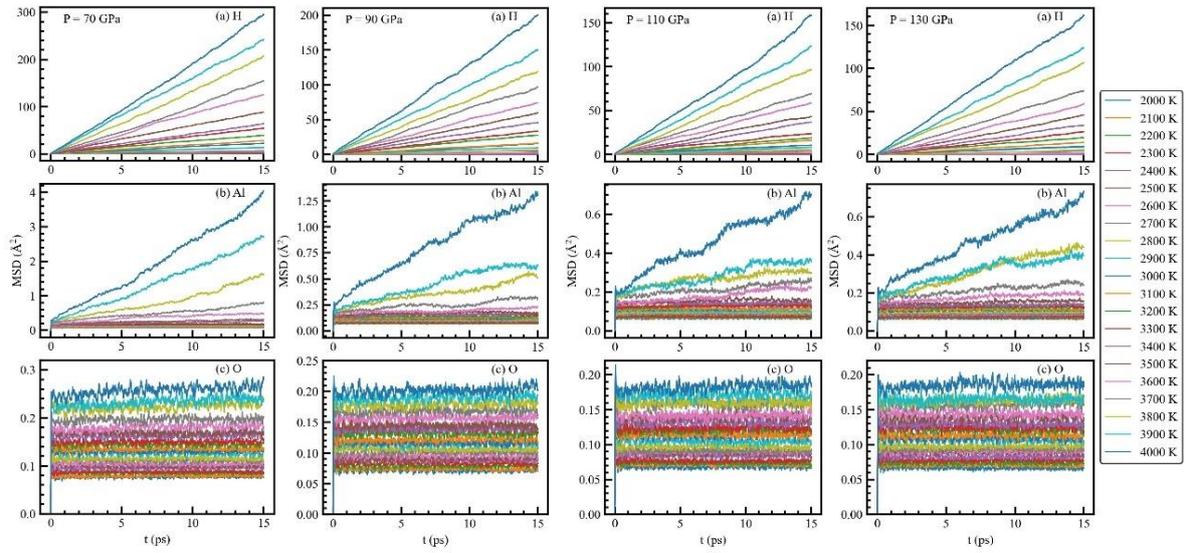

**Fig. S4.** The MSDs of H, Al, and O ions in δ-AlOOH at 70-130 GPa and 2000-4000 K.

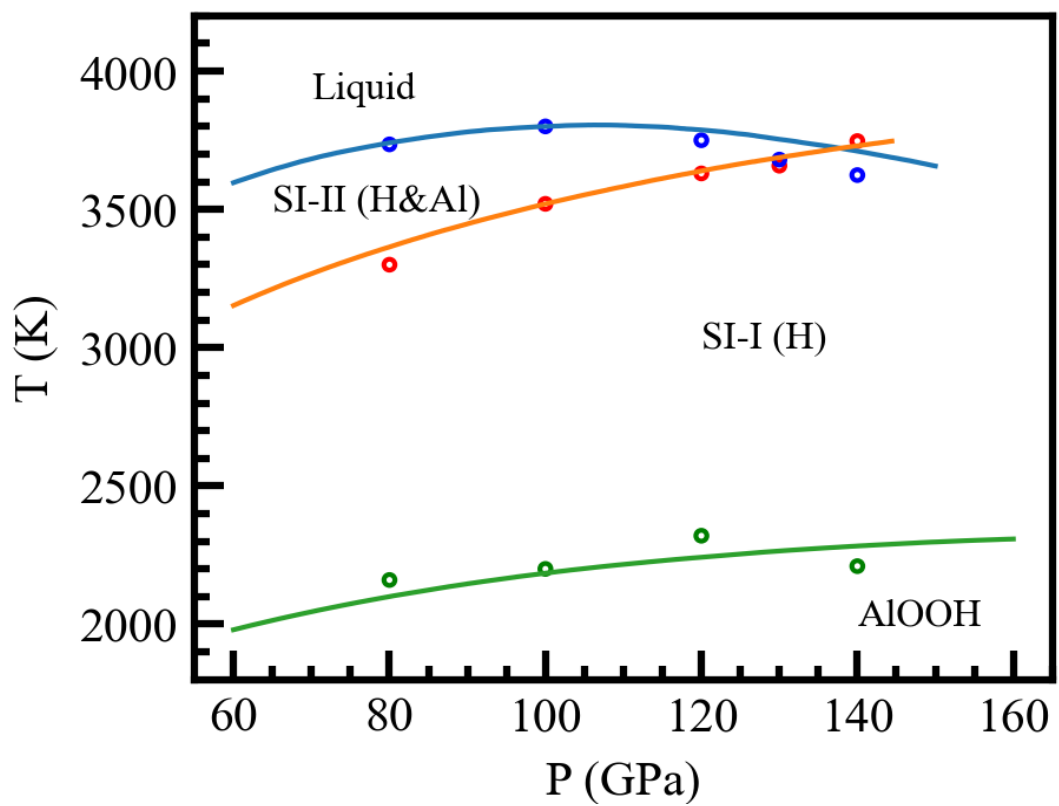

**Fig. S5.** The phase diagram of AlOOH at 60-160 GPa and 2000-4000 K.

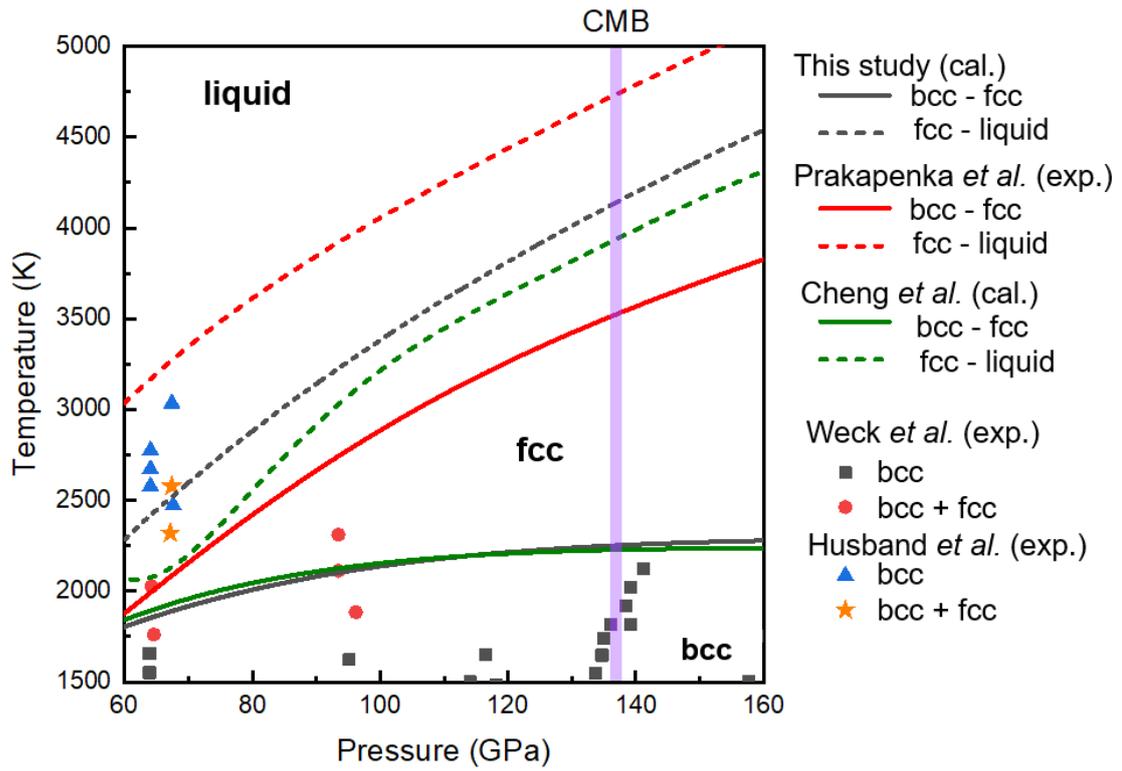

**Fig. S6.** Calculated phase diagram of ice at 60-160 GPa and 1500-4500 K using NeTI method in comparison with previous experimental[8-10] and computational[11] results. Thick purple line presents the pressure of core-mantle boundary (CMB).

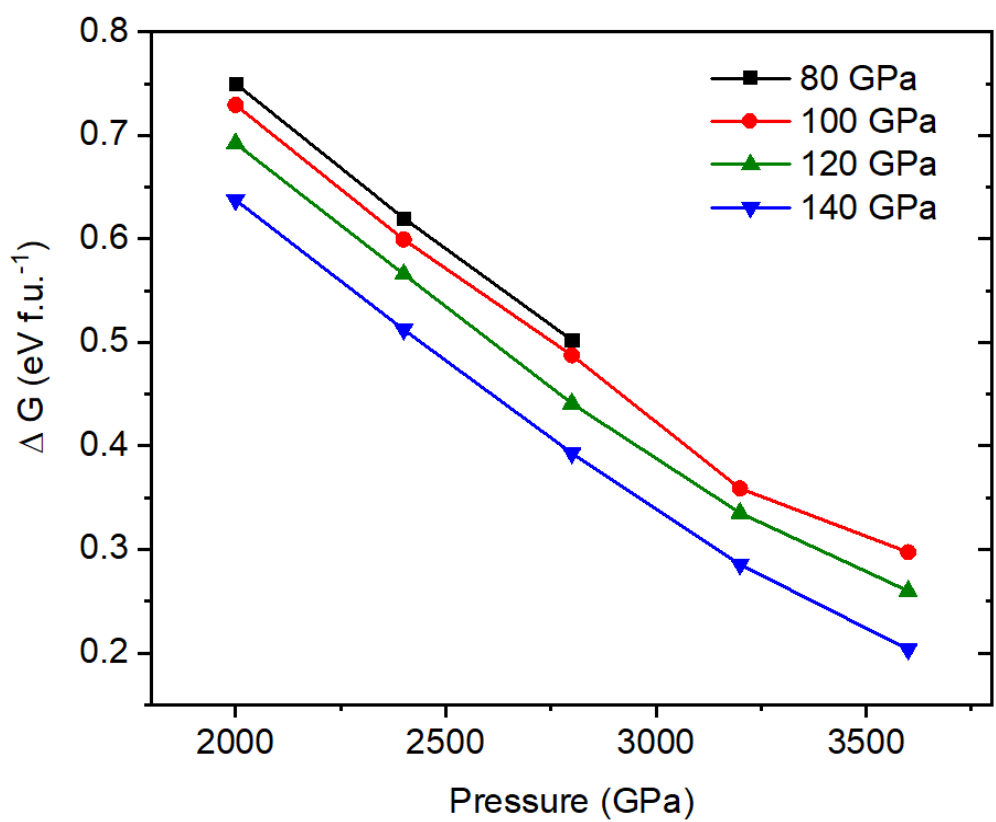

**Fig. S7.** Calculated Gibbs free energy differences (ΔG) of dehydration reaction of δ-AlOOH at 80-140 GPa and 2000-3600 K.

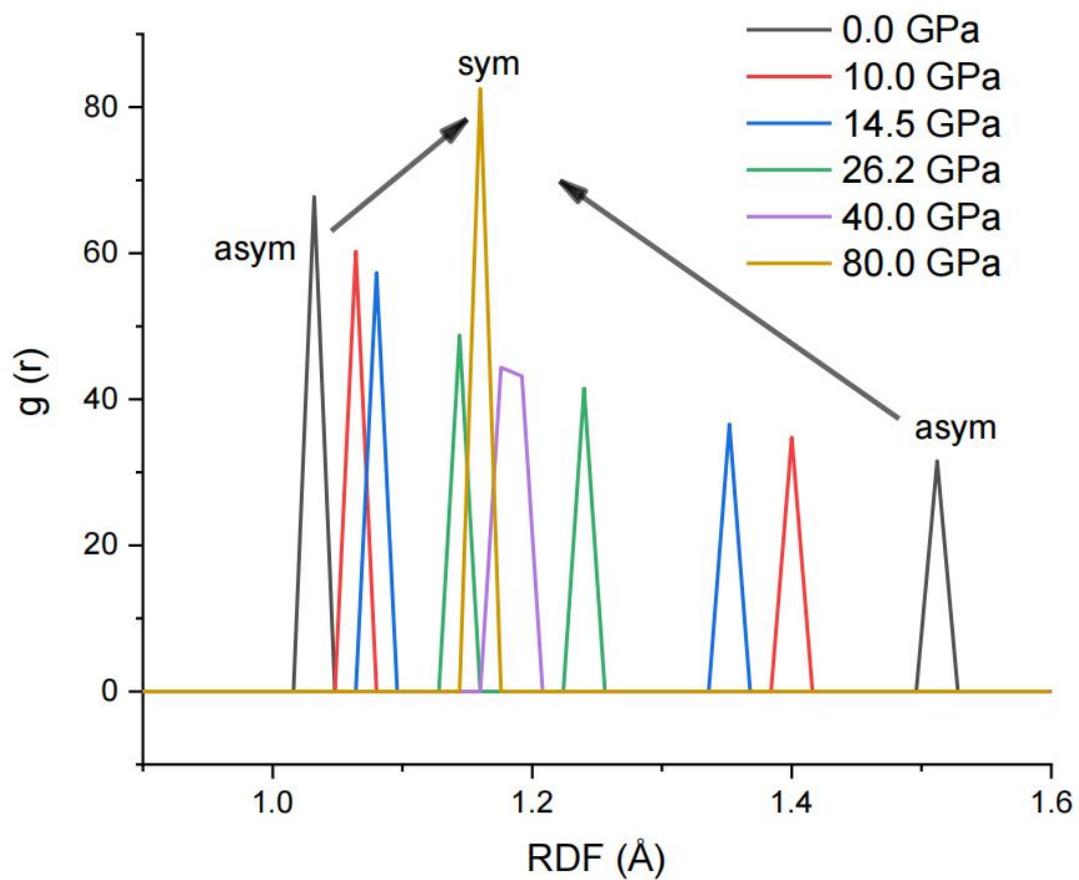

**Fig. S8.** Radial distribution functions (RDFs) of O-H in δ-AlOOH with increasing pressure indicting the asymmetric to symmetric O-H bonding transition at pressure ~40 GPa.

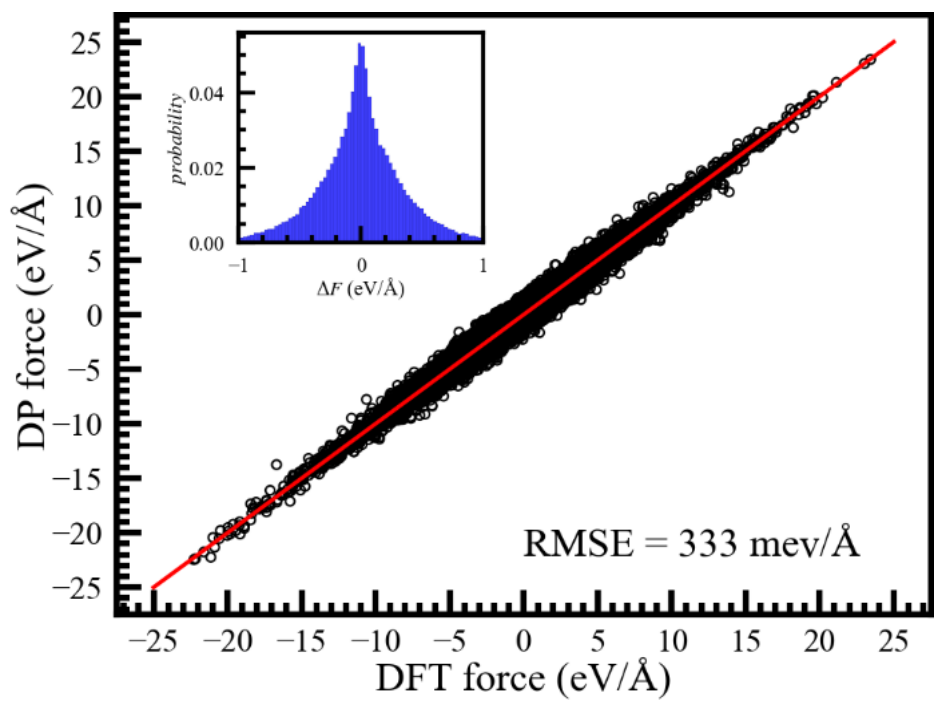

**Fig. S9.** Comparison of force predictions between the DP model and DFT calculations of configurations from the test data sets.

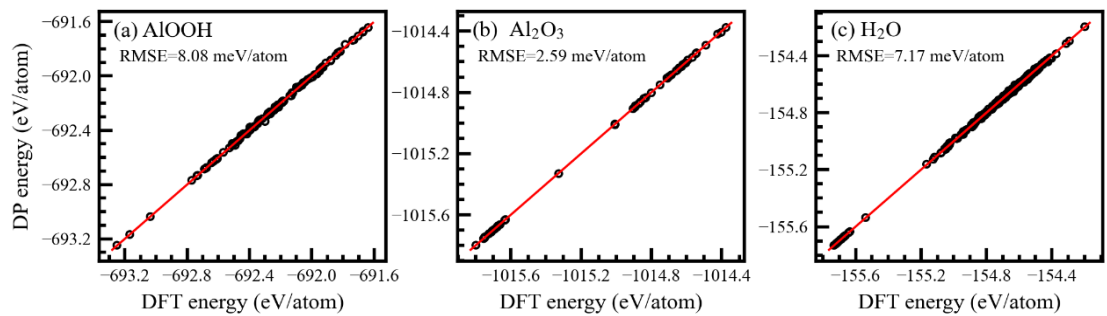

**Fig. S10.** Comparison of energy predictions between the DP model and DFT calculations of configurations from the test data sets.

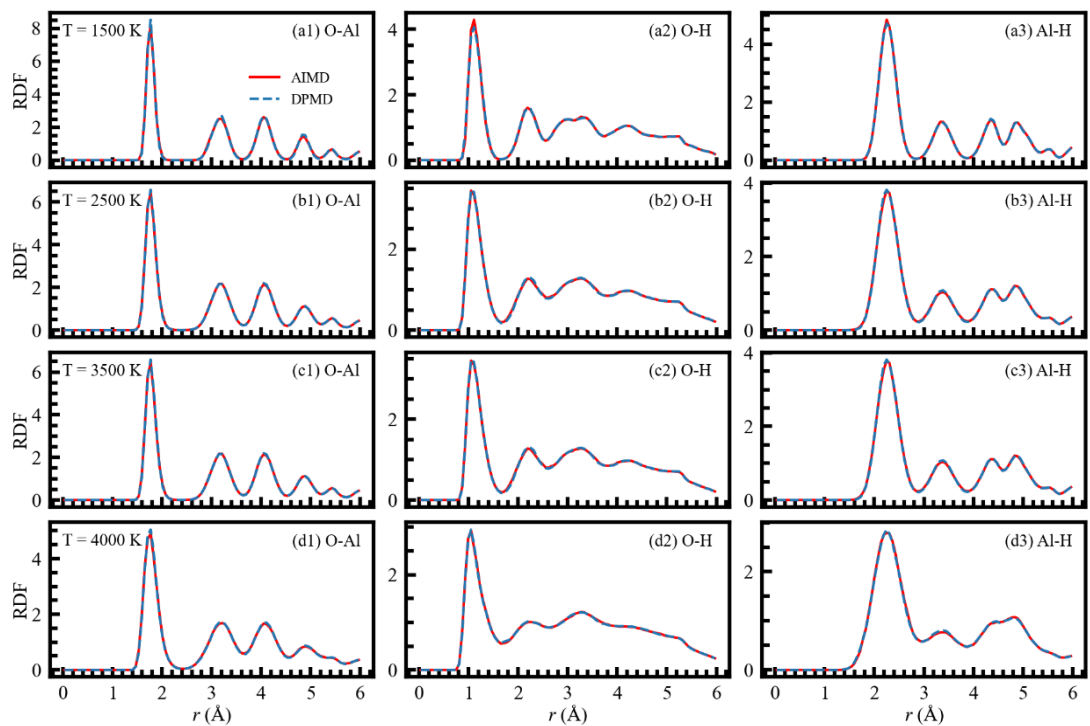

**Fig. S11.** Comparison of the RDFs obtained from AIMD and DPMD methods for AlOOH at 1500 K, 2500 K, 3500 K and 4000 K under 100 GPa.

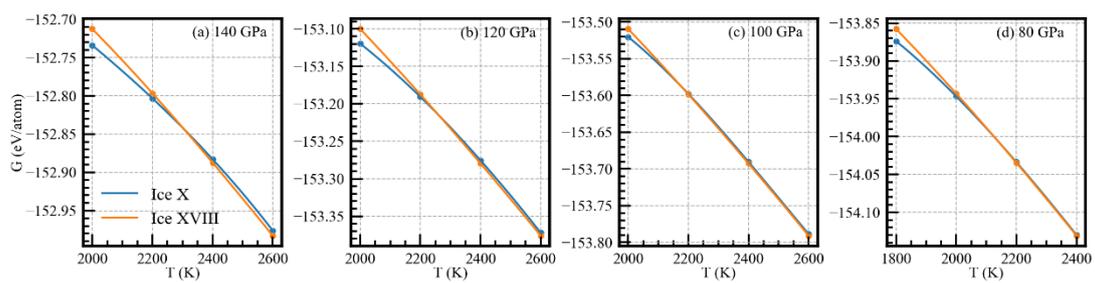

**Fig. S12.** The change Gibbs free energies of ice X (bcc) and ice XVIII (fcc, superionic) as a function of temperature at different pressures.

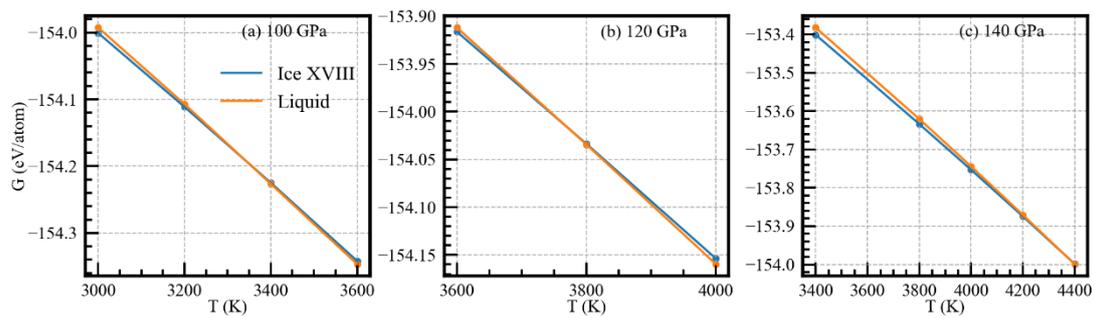

**Fig. S13.** The change of Gibbs free energies of superionic $H_2O$ ice XVIII and liquid $H_2O$ as a function of temperature at different pressures.

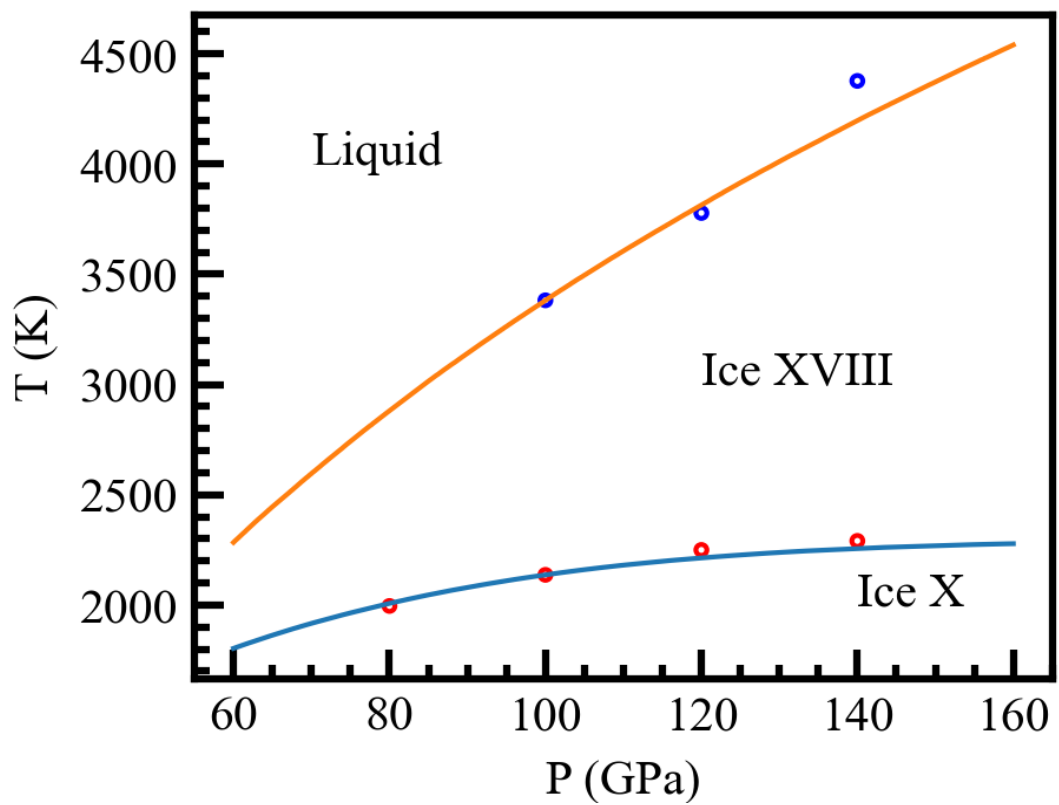

**Fig. S14.** The phase diagram of $H_2O$ ice at 60-160 GPa and 2000-4500 K. The phase transition boundaries between ice X (bcc structure) and ice XVIII (fcc structure) and liquid $H_2O$ are shown with blue and orange curves.

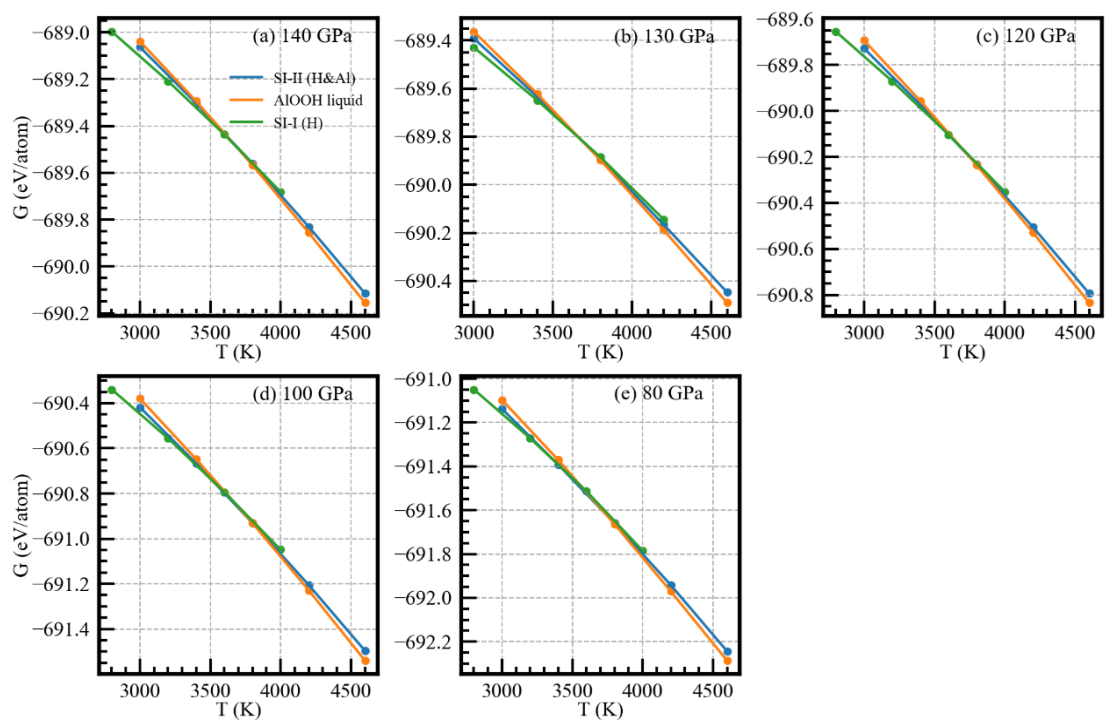

**Fig. S15.** The change of Gibbs free-energy of solid, SI-I, SI-II and liquid AlOOH as a function of temperature at 80-140 GPa.

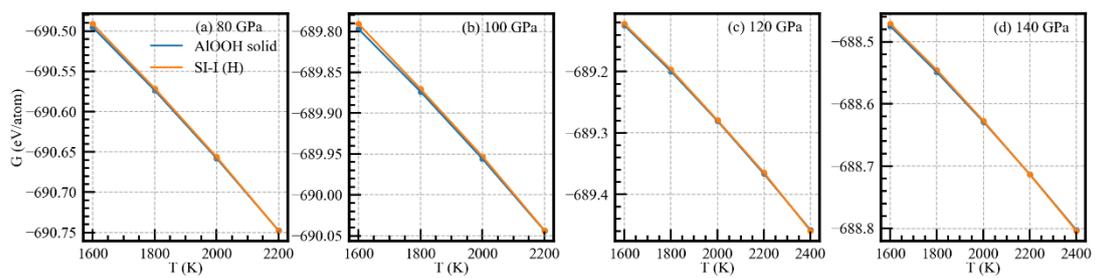

**Fig. S16.** The change of Gibbs free energies of solid and SI-I δ-AlOOH as a function of temperature at ~80-140 GPa.

Table S1. The transition points of water.

| P (GPa) | T (K) (ice X-ice XVIII) superionic water | T (K) (ice XVIII-liquid) water |
|---|---|---|
| 80 | 1996 | |
| 100 | 2137 | 3382 |
| 120 | 2250 | 3778 |
| 140 | 2291 | 4377 |

Table S2. The phase transition temperatures and pressures of AlOOH

| P (GPa) | T (K) SI-I→SI-II | T (K) SI-II→liquid | T (K) SI-I→liquid |
|---|---|---|---|
| 80 | 3300 | 3735 | 3580 |
| 100 | 3520 | 3800 | 3670 |
| 120 | 3630 | 3750 | 3685 |
| 130 | 3658 | 3680 | 3668 |
| 140 | 3747 | 3624 | 3689 |